\documentstyle[aps,twocolumn,epsf,graphicx]{revtex}

\newcommand{\beq}{\begin{equation}}
\newcommand{\eeq}{\end{equation}}        
\newcommand{\bqa}{\begin{eqnarray}}        
\newcommand{\eqa}{\end{eqnarray}}

\newcommand{\eq}[1]{{\frenchspacing Eq.~(\ref{#1})}}
\newcommand{\fig}[1]{{\frenchspacing Fig.~(\ref{#1})}}
\newcommand{\mpr}{\frac{m_{\pi}}{m{\rho}}}
\newcommand{\sesam}{{\sf SESAM}}
\newcommand{\txl}{{\sf T$\chi$L}}

\begin{document}

\draft

\title{ 
\hfill\begin{minipage}{0pt}\scriptsize \begin{tabbing}
\hspace*{\fill} IFUM 608-FT \\
\hspace*{\fill} IFUP-TH 10/98\\
\hspace*{\fill} HLRZ-98/6\\ 
\hspace*{\fill} HUB-EP-98/12\\
\hspace*{\fill} WUP-TH 98-6 \end{tabbing} 
\end{minipage}\\[8pt]  
Scanning the Topological Sectors of the QCD Vacuum with Hybrid Monte
Carlo}

\author{\frenchspacing B. All\'es}%
\address{ Dipartimento di Fisica, Universit\`a di Milano and INFN, Via
  Celoria 16, I-20133 Milano, Italy}

\author{\frenchspacing G. Bali} %
\address{Institut f\"ur Physik, Humboldt Universit\"at,
  Invalidenstrasse 110, D-10115 Berlin, Germany}

\author{\frenchspacing M. D'Elia} %
\address{Dipartimento di Fisica dell'Universit\`a and INFN, Piazza
  Torricelli 2, I-56126-Pisa, Italy, and Department of Natural
  Sciences, University of Cyprus P.O. Box 537, Nicosia CY-1678,
  Cyprus.}

\author{\frenchspacing A. Di Giacomo} %
\address{Dipartimento di Fisica dell'Universit\`a and INFN, Piazza
  Torricelli 2, I-56126-Pisa, Italy}

\author{N. Eicker,  K. Schilling, and A. Spitz}
\address{ HLRZ, c/o FZ-J\"{u}lich, and DESY-Hamburg, D-52425
  J\"{u}lich, Germany}

\author{\frenchspacing
S. G\"usken,   %
H. Hoeber,     %
Th. Lippert,   %
T. Struckmann, %
P. Ueberholz, and J. Viehoff}%
\address{ Fachbereich Physik, Universit\"at Wuppertal, D-42097
  Wuppertal, Germany}

\maketitle

\begin{abstract}
  We address a long standing issue and determine the decorrelation
  efficiency of the Hybrid Monte Carlo algorithm (HMC), for full QCD
  with Wilson fermions, with respect to vacuum topology.  On the basis
  of five state-of-the art QCD vacuum field ensembles (with 3000 to
  5000 trajectories each and $\mpr$-ratios in the regime $>0.56$, for
  two sea quark flavours) we are able to establish, for the first
  time, that HMC provides sufficient tunneling between the different
  topological sectors of QCD. This will have an important bearing on
  the prospect to determine, by lattice techniques, the topological
  susceptibility of the vacuum, and topology sensitive quantities like
  the spin content of the proton, or the $\eta '$ mass.


\end{abstract}

\section{Introduction}

In the past, a reliable access to topological quantities of the QCD
vacuum by methods of Lattice Gauge Theory has turned out to be  a real
challenge to the entire lattice approach.

Considerable progress has been achieved recently in the understanding
of how to extract topological observables which are inherently
continuum quantities from the discrete lattice.  It was
shown \cite{ALLESCRITICAL} that the field theoretical \cite{FABRICIUS}
and the geometrical \cite{LUESCHERTOP} definitions of the topological
charge yield---when suitably renormalized---equal values for the
topological susceptibility, $\chi$,  in the continuum. For practical
purposes, however, we  prefer the former definition, as it
offers superior quality of the stochastic signal \cite{ALLESCRITICAL}.

This being settled, there remains the severe problem: any trustworthy
stochastic sampling method must qualify to sufficiently decorrelate
the  members of the Monte Carlo time series with respect to the
observables of interest; and the folklore is that the topological
charge is the slowest of them all.

This nuisance is particularly aggravated in presence of dynamical
fermions in QCD as they induce long range interactions, in form of the
determinant of the fermionic matrix $M$ inside the path-integral.
Standard Hybrid Monte Carlo algorithms (HMC) deal with this non-local
problem by recourse to a stochastic Gaussian representation of
$\det(M)$ \cite{DUANE}. Such treatment implies the repeated {\em
  costly} solution of large systems of linear equations which puts
narrow limits on the affordable sample sizes, even on Teracomputers.
Thus autocorrelations on the HMC time series are an issue of prime
importance in judging the statistical reliability of estimates from
HMC in realistic settings.

A failure of stochastic methods to scan the topological sectors of the
QCD vacuum would be a real blow to lattice gauge theory, as topology
is known to be of great importance to elementary particle physics:
think of the Witten-Veneziano explanation of the $\eta '$ mass, the
relation of the axial vector current divergence to the topological
charge density through the Adler-Bell-Jackiw anomaly, or the r\^ole of
instantons in the structure of the vacuum.

For {\em staggered fermions}, indications have been given for quite
insufficient tunneling rates of the topological charge at very small
mass parameters, both for two and four flavors\cite{KURAMASHI,MMP}.

Some of us\cite{PISA} have performed simulations with four flavours of
staggered fermions using the HMC.  At $\beta=5.35$, on a $16^3\times
24$ lattice, the time history exhibits little mobility, over 450
trajectories if the quark mass is chosen as $ma=0.01$ corresponding to
$\mpr=0.57(2)$.  Much shorter autocorrelation times, of the order of
10 trajectories, are found at $ma=0.02$ and $ma=0.05$.  These quark
masses correspond to $\mpr=0.65(3)$ and $\mpr=0.75(1)$.

For full QCD with dynamical {\em Wilson fermions} trustworthy results
for the topological decorrelation efficiency of fermion algorithms
are missing in the literature.  In this note we report on an
investigation that will fill this gap.

Our results are based on the high statistics samples as obtained by
the \sesam\ \cite{MELBOURNE} and \txl\ \cite{STLOUIS} projects with
HMC on APE100 hardware in Italy and Germany.  The present sample
consists of three time histories on lattices of size $16^3\times 32$
\cite{TSUKUBA} at $\kappa=0.156$, $0.157$, and $0.1575$, corresponding
to three intermediate $\mpr$ ratios of $0.839(4)$, $0.755(7)$, and
$0.69(1)$, respectively. At the coupling $\beta=5.6$, the scale
computed from the $\rho$ mass is $a^{-1}=2.33(6)$ GeV
\cite{SESAMUPSILON}, after chiral extrapolation.  In the framework of
the \txl\ project, \sesam's set has been complemented by two ensembles
of configurations on $24^3\times 40$ lattices, again at $\beta=5.6$,
with $\kappa=0.1575$ and $0.158$, the latter value corresponding to
the lightest sea-quark mass attempted till now for Wilson fermions,
with $\mpr=0.56(2)$ \cite{TSUKUBA}.

So far, the \sesam\ sample has been partially analysed for integrated
autocorrelation times of standard observables only, like plaquette,
extended Wilson loops and octet hadron masses \cite{SESAMAUTOLATTICE}.
The autocorrelation time, $\tau _{int}$ of these observables appears to
be bounded by the one of the lowest eigenvalue of the Wilson fermion
matrix, the bound being of size 15 to 30 in units of molecular
dynamics time.  In this sense, we can start with sets of about 200
such `decorrelated' configurations to search for tunneling through
the topological sectors.

\section{Two Ways to Monitor for  Topological Charge}

The topological charge $Q$ is the integral over space-time of the
topological charge density, $Q(x)$, which is related via the $U_A(1)$
anomaly of the flavor singlet axial current,
$\partial^{\mu}J_{\mu}^{5}(x)=2N_f\, Q(x)$.

To monitor for the topological content along the time histories, we
use two independent methods to estimate $Q$:
\begin{enumerate}
\item The {\em gluonic} calculation, which starts from the charge
  density
\begin{equation}
  Q(x) =
  \frac{g^2}{64\pi^2}\epsilon^{\mu\nu\rho\sigma}F^{a}_{\mu\nu}(x)F^{a}_{\rho\sigma}(x).
\end{equation}
For the lattice definition of $Q$ we follow Ref.~\cite{FABRICIUS}.
Local fluctuations of the gauge fields are suppressed by the standard
cooling procedure \cite{TEPER} just as in Ref.~\cite{PISA}.
\item The {\em fermionic} evaluation follows the Smit and Vink
  proposal \cite{SMIT} which is inspired by the Atiyah-Singer theorem:
\begin{equation} \label{smitvinck}
  Q = m\, \kappa_P \langle Tr(\gamma_5 G)\rangle_U.
\end{equation}
Here $G$ represents the quark propagator in the background gauge field
$U$ with quark mass $m$.  $\kappa_P$ is a renormalization constant.
No cooling is applied here.
\end{enumerate}
We comment that both methods are in principle prone to characteristic
systematic errors: while the gluonic approach might suffer from loss
of instantons due to cooling, the fermionic monitor might be affected
by 'dislocations'. It would therefore be gratifying to see them agree in
their monitoring functionality.

To our knowledge, the Smit-Vink proposal has never been really
exploited.  Computing the trace in \eq{smitvinck} requires the
application of stochastic estimator techniques.  Such estimates in
general turn out to be extremely noisy.  However, thanks to recent
progress in noise reduction methods achieved by some of us
\cite{VIEHOFF}, we are now in the position to extract rather accurate
signals for $\mbox{Tr} (\gamma_5 G)$.

\section{Results}
With the procedures decribed in the previous section we are
sufficiently equipped to monitor for the decorrelation of the
topological charge in the HMC process.
 
\begin{figure}
\centerline{\includegraphics[width=\columnwidth]{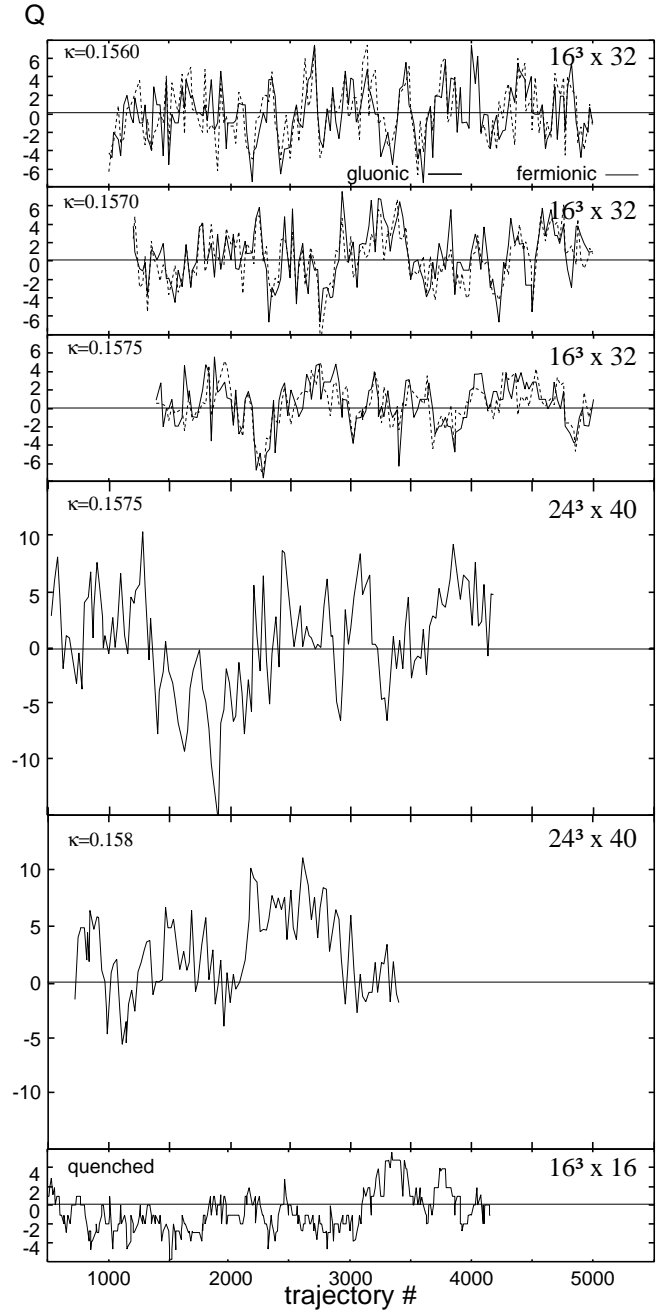}}
\vspace*{8pt}
\caption{
  Time series of topological charge. We plot series from 3 \sesam\ and
  2 \txl\ ensembles. The last canvas is a quenched HMC simulation at
  $\beta=6.0$.
\label{FIG:5SERIES}}
\end{figure}

\fig{FIG:5SERIES} shows the time series for $Q$ as measured on our
ensembles. The first three graphs contain 200 configurations each.
These ensembles were taken from contiguous series of trajectories of
length 5000, produced by \sesam\ on $16^3\times 32$ lattices. The
configurations are separated by 25 units in absolute molecular
dynamics time.  In the first three graphs, the results from the
gluonic and fermionic measurements are super-imposed.  We find them to
agree nicely which demonstrates that the systematic errors of both
methods are sufficiently under control.

First inspection shows many tunneling events of the topological charge
between the distinct sectors, for all three sea quark masses on the
$16^3\times 32$ lattices. There seems to be a tendency that the
fluctuations decrease with increasing $\kappa$.  In an attempt to
quantify these features we introduce the mobility of tunneling, $D_d$,
as the average weighted number of tunneling events measured on
configurations which are separated by a distance of $d$ along the time
history:
\begin{equation}
D_d = \frac{1}{N-1}\sum_{i=1}^{N-1}|(\tilde Q(i+1)- \tilde Q(i))|,
\label{MOBILITY}
\end{equation}
with $\tilde Q$ being the integer value of the measured quantity $Q$
and $N$  the number of measurements.  The resulting numbers are
given in table~\ref{TAB:FREQUENCY}.  They reflect the tunneling rates
and thus the decorrelation capability of HMC, when approaching the
chiral limit. 

It should be noted, however, that $D_d$ depends on the volume for two
reasons: $D_d$ has not been normalised to the amplitude of the
fluctuations and our HMC parameters have been retuned when the system
size was changed.

\begin{figure}
\centerline{\includegraphics[width=\columnwidth]{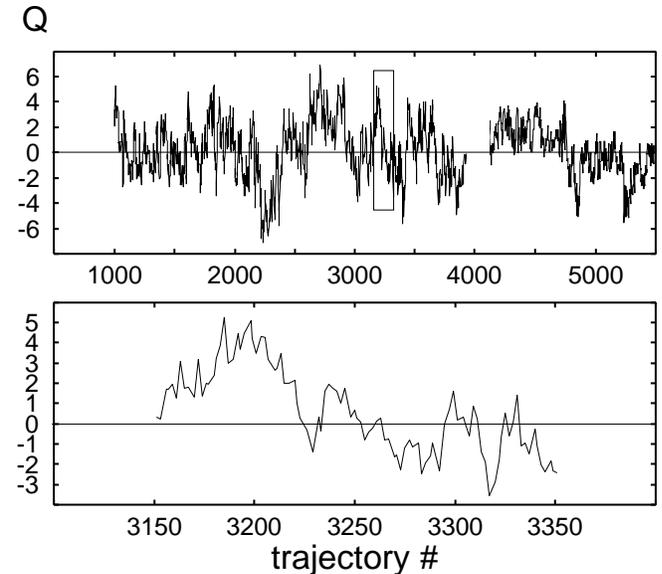}}
\caption{Fine scan of 
  of $Q$ for $\kappa=0.1575$ on the $16^3\times 32$ lattice.  The
  second canvas is a zoom.
\label{FIG:FINE}}
\end{figure}

With the entire time histories of our HMC runs being archived we are
able to postprocess all trajectories from the \sesam\ and \txl\ 
samples. For $\kappa=0.1575$---the smallest \sesam\ mass---we have
computed the topological charge on a contiguous segment of $> 5000$
trajectories for every second trajectory, see \fig{FIG:FINE}.  We find
the topological charge to change frequently on that time scale.  This
is illustrated in the zoom of \fig{FIG:FINE}, which shows considerable
fluctuations on that time scale.

\begin{figure}
\centerline{\includegraphics[width=\columnwidth]{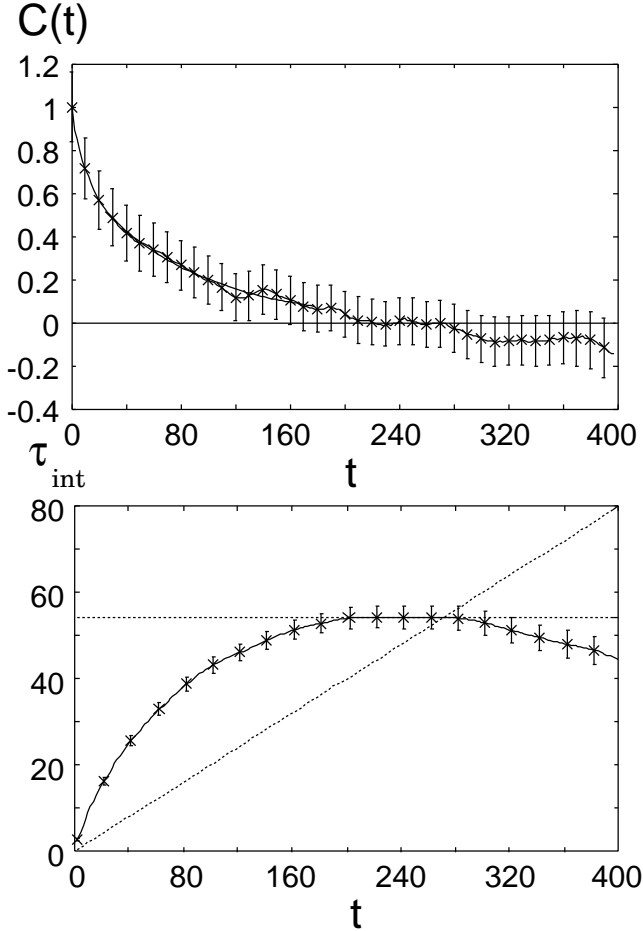}}
\caption{Autocorrelation function and integrated autocorrelation time 
  of $Q$ for $\kappa=0.1575$ on the $16^3\times 32$
  system.\label{FIG:AUTO}}
\end{figure}

The series for $\kappa=0.1575$ is long enough to allow for the
computation of the autocorrelation function, as plotted in
\fig{FIG:AUTO}.  We can apply an exponential fit to its `slowest'
mode.  The fit reveals the exponential autocorrelation time of
$\tau_{exp}=80(10)$.  The integrated autocorrelation time from the
lower diagram in \fig{FIG:AUTO} turns out to be $\tau_{int}=54(4)$.
These numbers should be compared to the upper bound of $\tau_{int}$ as
estimated from other observables,  
$\max\tau_{int}=42(4)$ \cite{SESAMAUTOLATTICE}.

\begin{figure}
\includegraphics[width=.95\columnwidth]{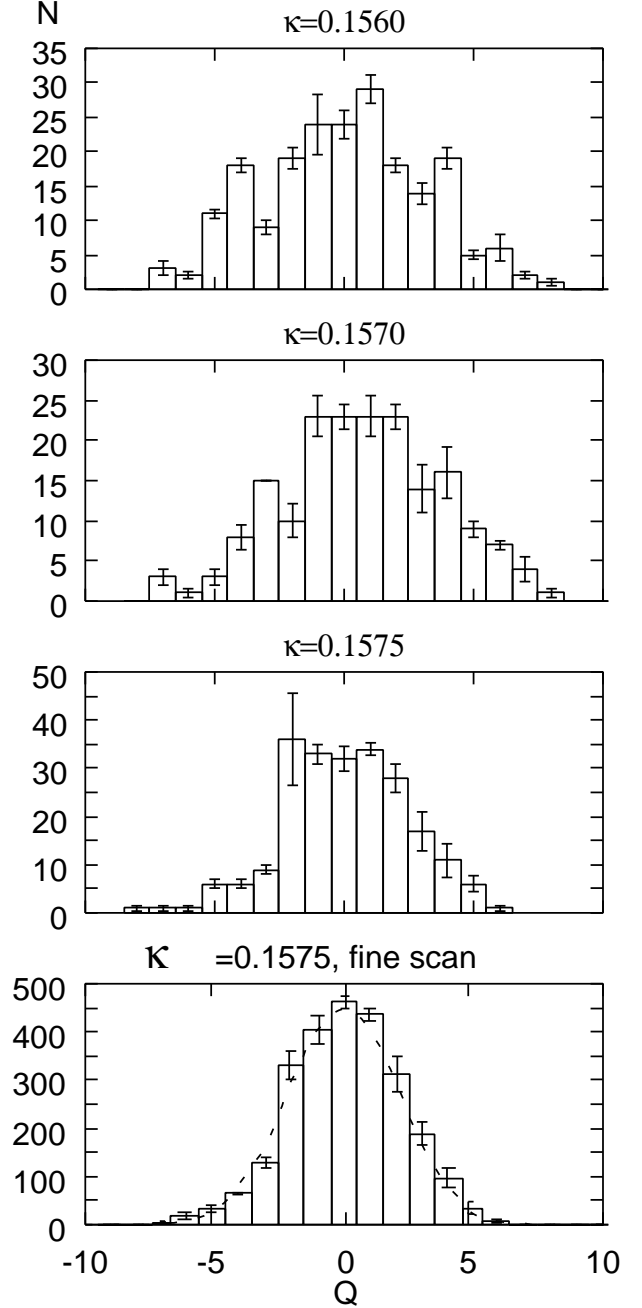}
\caption{Histograms of the topological charge.
\label{FIG:4HISTOS}}
\end{figure}

The tunneling behaviour is reflected in the histograms of
\fig{FIG:4HISTOS}.  The errors quoted have been estimated by naive
jackknife. The distributions are well peaked at $Q=0$. With 200
entries, though, the symmetry is not yet so well established.

Histogramming the fine scan of \fig{FIG:FINE}, we arrive at the fourth
histogram which comprises 2500 entries.  We find a smooth Gaussian
shape to emerge, see \fig{FIG:4HISTOS}.

Finally, we turn to the volume dependency. On the $24^3\times 40$
system, at $\kappa=0.1575$, the time history looks rather much alike
to the one on the $16^3\times 32$ lattice, apart from an apparent
increase in amplitude.  The mobility at a separation of 25 units of
molecular dynamics time, $D_{25}$, turns out to be twice as large as
for the small system, see table~\ref{TAB:FREQUENCY}.  Naively, we
would expect that the mobility scales approximately with the square
root of the volume, leading to a factor of two.  However, we remark
that on the large system, we had to re-adjust the length of the
molecular dynamics trajectory to $T=0.5$.  Thus, for the chosen
separation of $T=25$, we refresh momenta and fermions twice as often
as on the smaller system.

Going more chiral, at $\kappa=0.1580$, our present sample size is not
sufficient to make definite statements, as the expected symmetry
around $\langle Q\rangle=0$ is not yet established.

The last entry in \fig{FIG:5SERIES} refers to a HMC series in the
quenched case at $\beta=6.0$, on a $16^4$ lattice.  The reduced volume
amounts to a smaller amplitude.  The series appears to be less ergodic
on the time scale of 4000 trajectories, compared to the three full-QCD
runs on the $16^3\times 32$ lattice. The mobility in the quenched
case is observed to take the value $1.2$.

\section{Discussion and Outlook}

We find that HMC is capable to create sufficient tunneling between the
vacuum sectors when dealing with dynamical Wilson fermions for
$\mpr>0.56$. 
This is seen with both, gluonic and fermionic
determinations of $Q$ which proves their reliability as monitoring
devices.

\begin{table}
\caption{Mobility  $D_{25}$, see eq.~\ref{MOBILITY}.
\label{TAB:FREQUENCY}}
\begin{tabular}{lcccc}
  $\kappa$ & 0.1560 & 0.1570 & 0.1575 &
  0.1580 \\
  \hline
  $16^3\times 32$ & 2.8  & 2.5 & 1.9 &  -    \\
  \hline
  $24^3\times 40$ & -    & -   & 3.8 & 2.8    \\
\end{tabular}
\end{table}

The autocorrelation rates as determined from various non-topological
quantities are in accord with the autocorrelation time of the
topological charge. It is interesting to note that the high frequency
mobility of HMC with respect to topology is important for achieving
the expected symmetry of the topological charge distributions.

When approaching the chiral regime, at $\kappa = 0.1580$ or
$\mpr=0.56$, we would expect to need a sample size of order $10.000$
to achieve sufficient ergodicity with respect to topology. This is
definitely in the reach of teracomputing.

It is evident that the fermionic force plays an important r\^ole for
the tunneling efficiency of HMC: in the quenched setting, HMC appears
to be less efficient in driving the system through the topological
sectors.  In order to understand this effect in more detail further
studies are needed.

At this point, with the available sample, we are in the position to
study hadronic properties related to topology, like the proton spin
content and the topological susceptibility for $\mpr$-values down to
$0.69$.  Work along this line is in progress.

Another line of research being followed by some of us \cite{CYPERN} is
the tunneling efficiency of alternative stochastic algorithms, such as
the multibosonic method \cite{LUESCHER}.

\acknowledgements{ We are most grateful to the Zentralinstitut f\"ur
  Angewandte Mathematik at FZ-J\"ulich for allocating three terabyte
  of archive space to the \sesam-\txl-projects, which was
  indispensible for this work. We thank the INFN group at La Sapienza,
  Roma, in particular Dr.\ F.\ Rapuano, for support.
  
  We acknowledge support by the DFG (grants Schi 257/1-4, Schi
  257/3-2, Ba 1564/3-1 and Ba 1564/3-2) and the Wuppertal
  DFG-Graduiertenkolleg ``Feldtheoretische und Numerische Methoden in
  der Statistischen und Elementarteilchenphysik''.  The Pisa group
  acknowledges support by MURST and by EC, contract FMRX-CT97-0122.  }


\begin{thebibliography}{99}
\frenchspacing
%
\bibitem{ALLESCRITICAL} B. All\'es, M. D'Elia, A.  Di Giacomo, and R.
  Kirchner, {\em A Critical Comparison of Different Definitions of
    Topological Charge on the Lattice}, hep-lat/9711026.
%
\bibitem{FABRICIUS} M. Campostrini, A.  Di Giacomo, and H.
  Panagopoulos, Phys. Lett. {\bf B212} (1988) 206; B. All\'es, M.
  Campostrini, A.  Di Giacomo, Y.  G\"und\"uc, and E.  Vicari, Phys.
  Rev. {\bf D 48} (1993) 2284.
%
\bibitem{LUESCHERTOP} M. L\"uscher, Comm. Math. Phys. {\bf 85} (1982)
  39.
%
\bibitem{DUANE} S. Duane, A. Kennedy, B. Pendleton, and D. Roweth, Phys.
  Lett. {\bf B195} (1987) 216.
%
\bibitem{KURAMASHI} Y. Kuramashi et al., Phys. Lett. {\bf B313} (1993)
  425.
\bibitem{MMP}{M.~Mueller-Preussker, in Proc. of XXVI Int. Conf. on High
Energy Physics, Dallas 1992, Ed. J.R.~Sanford, AIP Conference
Proceedings
No. 272, 1545 (1993)}
%
\bibitem{TEPER}
B. Berg,  Phys. Lett. {\bf B104} (1981) 475.\\
M. Teper, Phys. Lett. {\bf B162} (1985) 157.
%
\bibitem{PISA} B.~All\'es, G.~Boyd, M.~D'Elia, A.~Di~Giacomo, and
    E.Vicari, Phys. Lett. {\bf B389} (1996) 107.  
%
\bibitem{SMIT} J. Smit and J. C. Vink, Nucl. Phys. {\bf B286} (1987)
    485.
%
\bibitem{MELBOURNE} SESAM-Collaboration: U. Gl\"assner, S. G\"usken,
    H. Hoeber, Th.\ Lippert, X. Luo, G.  Ritzenh\"ofer K.\ Schilling
    and G. Siegert, in T. D. Kieu, B. H.  J. McKellar, and A. J.
    Guttmann, (edts.): Proceedings of {\it Lattice '95}, Nucl.\ Phys.\ 
    {\bf B} (Proc.\ Suppl.\ ) {\bf 47} (1996) 386-393.
%
\bibitem{STLOUIS} T$\chi$L-Collaboration: L. Conti, N. Eicker, L.
    Giusti, U.  Gl\"assner, S. G\"usken, H.  Hoeber, Th.\ Lippert, G.
    Martinelli, F.  Rapuano, G.  Ritzen\-h\"o\-fer, K.\ Schilling, G.
    Siegert, A. Spitz, and J. Viehoff, Nucl.\ Phys.\ {\bf B} (Proc.\ 
    Suppl.\ ) {\bf 53} (1997) 222.
%
\bibitem{TSUKUBA} Th. Lippert, G. Bali, N. Eicker, L. Giusti, U.
  Gl\"assner S. G\"usken, H. Hoeber, P. Lacock, G. Martinelli, F.
  Rapuano, G.  Ritzenh\"ofer, K. Schilling, G. Siegert, A. Spitz P.
  Ueberholz, and J. Viehoff, Nucl. Phys. {\bf B} Proc. Suppl. {\bf 60A} (1998)
  311.
%
\bibitem{SESAMUPSILON} N. Eicker, Th. Lippert, K. Schilling, A. Spitz,
  J. Fingberg, S. G\"usken, H. Hoeber, and J. Viehoff, {\em Improved
    Upsilon Spectrum With Dynamical Wilson Fermions}, HLRZ-1997-35,
  hep-lat/9709002.
%
\bibitem{SESAMAUTOLATTICE} Th. Lippert et al., {\em Critical Dynamics
    of the Hybrid Monte Carlo Algorithm}, HLRZ-1997-73,
  hep-lat/9712020.
%
\bibitem{VIEHOFF} \sesam-collaboration, to be published.
%
\bibitem{CYPERN} C. Alexandrou, M. D'Elia, A.  Di Giacomo, and H.
  Panagopoulos, in preparation.  \nonfrenchspacing
%
\bibitem{LUESCHER} M. L\"uscher, Nucl. Phys. {\bf B418} (1994) 637;\\
  A. Borrelli, P. deForcrand, and A. Galli, Nucl. Phys. {\bf B477} (1996)
  809.
%
\end{thebibliography}
\end{document}